\documentclass[11pt,letterpaper]{article}
\usepackage[top=0.85in,left=0.75in,footskip=0.75in]{geometry}
\usepackage[utf8]{inputenc}
\usepackage{authblk}

\usepackage{graphicx}
\usepackage{amsmath}
\title{ K-clique percolation  in free association networks. The mechanism 
behind the  $7 \pm 2 $ law ?}
\author[1,*]{O. Valba}
\author[2]{A. Gorsky}
\affil[1]{Department of Applied Mathematics, MIEM, National Research University Higher School of Economics, Moscow, Russia}
\affil[2]{Kharkevich Institute for Information Transmission Problems RAS,  Moscow, Russia}
\affil[*] {ovalba@hse.ru}
\date{}                     
\makeatletter
\renewcommand\@biblabel[1]{#1.}
\makeatother
\begin{document}
\maketitle
\begin{abstract}
It is important to reveal the mechanisms of propagation
in  different cognitive networks.
In this study we discuss the k-clique percolation phenomenon on
the free association networks including  "English Small World of Words project" (SWOW-EN). 
We compare different semantic networks and networks of free associations for different languages. 
Surprisingly it turned out that  $k$-clique percolation
for all $k<k_c=(6-7)$ is possible on  SWOW-EN and Dutch language network.
Our analysis suggests the new universality patterns for a community organization of free  association networks.
We conjecture that our result can provide the qualitative explanation of the 
Miller's $7\pm 2$ rule for the capacity limit of  working memory. The
new model of network evolution extending the preferential attachment is suggested  which provides the observed value of $k_c$.

\end{abstract}
\section{Introduction}

Networks represent powerful models for exploring different cognitive systems and processes \cite{siew2019,rev2}. For example, in \cite{stella2018,stella2019} the authors propose multiplex network model of  the formation of mental lexicon and early word acquisition. In \cite{kenett2014} the structural properties of semantic networks for low and high creativity people are discussed. In \cite{smith2013,bourgin2014,olteteanu2015,olteteanu2017,valba2021} network-based methods are used for simulation the  mechanisms of solving Remote Associates Tests, allowing to estimate a human's creative potential \cite{mednik1962}. 

Complex networks often exhibit meso-scale or global characteristics of structural order.Certain networks exhibit community structure, in which densely connected communities of nodes exhibit sparse or weak inter-community connections. In semantic networks, one word can belong to several communities, so standard community detection methods are not applicable.
We investigate the community organization of the free association network focusing at one  described in \cite{dedeyne2019}, known as "English Small World of Words project" (SWOW-EN). This network differs from other datasets in a higher density, which is achieved by the presence of links of weak association strength. The dense network structure allows us to study  k-clique community organization of larger k. We compare its properties with different semantic networks of the English and Dutch languages and networks of free associations. 

This study is mainly focused  at the percolation analysis of the free association
networks. The percolation approach was used to quantify the 
flexibility of one or another network characteristics
of semantic network \cite{ kenett2018, stella2020,arenas2011}. 
In \cite{kenett2018} flexibility of thought is investigated by  percolation analysis and the cognitive declines due to aging have been discussed.
In context of the creativity theory the percolation analysis has been discussed
in \cite{kenett2014,kenett2018,stella2020} and it was demonstrated that the semantic network of the high-creative group broke apart slower that that of less-creative group.
It was also shown via percolation approach  in \cite{stella2020} that across the lifespan the mental
lexicon is fragile against the combined semantic and phonological attacks.

More general  phenomenon  involves the percolation 
of k-clique introduced in \cite{derenyi2005,palla2007}. For the random Erdos-Renyi
ensemble the critical link probability for any k can be found
analytically however for  real networks an estimation of the
critical threshold for k-clique percolation is a nontrivial problem.
In the cognitive networks the k-clique percolation has been
recently discussed in \cite{kenett21} for the problem of 
aging in the semantic memory.

In this study we investigate a k-clique percolation in the 
free association networks and question if there is some
upper bound $k_c$ when  no k-clique percolation exists
for $k>k_c$. A bit surprisingly it turned out that 
there is the sharp bound at $k_c=6-7$ both for
English and Dutch languages.

The sharp bound for clique percolation certainly provides
the information concerning the structural organization 
of  free association networks. However it certainly also influences 
the effectiveness
of the processes on the network since  k-clique
percolation is a particular  dynamical process. Only the k-cliques 
with $k<k_c$ can propagate effectively 
through the free association network. The discussion concerning
the distinction between a structure and a process 
in  semantic networks can be found in 
\cite{dis1,dis2}.
 
 The test protocols  for free associations allow very 
 short time intervals for answering
 hence we can consider them as a kind of probe of 
 working memory. On the other hand the limitation
 of the working memory capacity is well-known phenomenon
 \cite{miller,cowan} and the person can remember simultaneously
 only a finite number of items of different nature although
 there is some mild dependence on a nature of item. This
 phenomenon is known as Miller's $7\pm 2$ law. We conjecture
 that our finding could serve as potential explanation of the 
 mechanism behind the Miller's law. Indeed we have to remember
 the k-linked items for some short period of time. This 
 can be considered as the k-clique percolation process in 
 some effective "working memory network".
 
 Looking at the mechanism responsible for the limit 
 of working memory capacity  
 the natural question concerns the evolutionary origin of the 
 particular value of $k_c$ and the rules of evolution
 which bring the network to this particular value of $k_c$.
 We suggest new rule of network evolution which can be considered
 as the modification of preferential attachment when a new 
 node it attached to both nodes connected by link. It turns
 out that this new mechanism provides the desired value
 $k_c=(5-6)$ for different sizes of the network.

\section{Data description}

The free association network SWOW-EN is a weighted directed network with $N=12~217$ stimuli words. Stimulus materials (cue words) were constructed using a snowball
sampling method, allowing authors \cite{dedeyne2019} to include both frequent and less frequent cues at the same time. The final set consists of 12~292 cues (stimuli), the weight of the link indicates fraction of the experiment participants which gave this particular response to a cue (i.e. the conditional probability of a response given a cue). Therefore, the total weight of links going out of each node is less or equal to 1. For our analysis we consider the network as undirected, attributing the greatest weight to an edge in the case of a bidirectional association. 

Also we analyze the free association network, based on the  South Florida free association data base \cite{nelson2004} and the  free association network, known as the Edinburgh Associative Thesaurus \cite{edinburgh}. 

Besides, we use networks, containing taxonomic relations (e.g. “A is a type of B”), synonymous relation (e.g. “A also means B”) and phonological similarities. All data were retrieved  from Wolfram Research \cite{worddata}, which mostly coincides with WordNet 3.0 \cite{miler}. Finally, we study free association networks for Russian and  Dutch languages. We used Russian thesaurus \cite{rus_th} and Dutch association data \cite{swow}, removing words, that have no associations. 
Table ~\ref{tab:1} summarize the basic structural properties of used networks.
\begin{table}[ht]
\caption{Structural properties of semantic networks.}
\begin{center}
\begin{tabular}{l|l|l|l|l|l|l|l}
\hline \hline
Network & Nodes\hspace{1cm} & Edges & Density & Transitivity & Clustering &$p_c(2)$& $p_c(3)$ \\ \hline\hline
SWOW-EN free association& 12~217& 352~403&0.0047 &0.052&0.113&$8.2\cdot 10^{-5}$& 0.0064\\ \hline
Florida free association& 5~019&55~246&0.0044 &0.083&0.186&$2.0\cdot10^{-4}$&0.0100 \\ \hline
Edinburgh free association &6~437&36~921&0.0017&0.059&0.124&$1.6\cdot10^{-4}$&0.0088\\ \hline \hline
Taxonomic &7~943&42~042&0.0013&0.048&0.093&$1.3\cdot10^{-4}$&0.0079\\ \hline
Synonyms &6~526&13~134&0.0006 &0.284&0.344&$1.5\cdot10^{-4}$&0.0088\\ \hline
Phonological &4~618&15~447&0.0014&0.345&0.246&$2.2\cdot10^{-4}$&0.0104 \\ \hline
Multiplex &8~383&68~505&0.0019.&0.112&0.283&$1.2\cdot10^{-4}$&0.0078 
\\ \hline \hline
RUS thesaurus &5~377&51~191&0.002 &0.067&0.163&$1.9\cdot10^{-4}$&0.0096\\ \hline
Dutch Data &10~486&207~810&0.0038 &0.067&0.163&$9.5\cdot10^{-5}$&0.0069\\ \hline\hline

\end{tabular}
\end{center}
\label{tab:1}
\end{table}
\section{K-clique percolation}
We begin with a few definitions laying down the fundamentals of k-clique
percolation \cite{derenyi2005,palla2007}. \emph{K-clique} is a complete (fully connected) subgraph of k vertices.
We say, that two k-cliques are \emph{adjacent} if they share $k-1$
vertices, i.e., if they differ only in a single vertex. A subgraph, which is the union of a sequence of adjacent k-cliques is called \emph{k-clique chain} and two k-cliques are \emph{k-clique-connected}, if there
exists at least one k-clique chain containing the two k-cliques. Finally, \emph{k-clique percolation cluster} is defined as a maximal k-clique-connected subgraph, i.e., it is the union of all k-cliques that are k-clique-connected to a particular k-clique.

The Erdosh–Renyi random graphs show a series of interesting transitions when the probability $p$ of two nodes being connected is increased. For $k=2$ the transition is well known and manifested by the appearance of a giant component in a network at critical probability $p_c(k=2)=\frac{1}{N}$, where $N$ is the number of nodes. For each $k$ one can find a certain threshold probability $p_c(k)$ above which the k-cliques organize into a giant community \cite{palla2007}:
$$
p_c(k)=\frac{1}{\left[ N(k-1)\right]^{\frac{1}{k-1}}}.
$$
Table ~\ref{tab:1} contains the values $p_c(k)$ with $k=2,3$ for random networks of the same size as semantic networks. We found, that network density,i.e. the observed link probability, for all datasets satisfy the inequality $p_c(2)<\rho<p_c(3)$. That is, if links in a semantic network were formed randomly, then all the vertices are included in the percolation cluster of $ k = 2 $, that is,one connected component, but do not form a cluster of $ k = 3$. 

\section{K-clique community organization of semantic networks}
\begin{figure}[ht]
\centerline{\includegraphics[width=20cm]{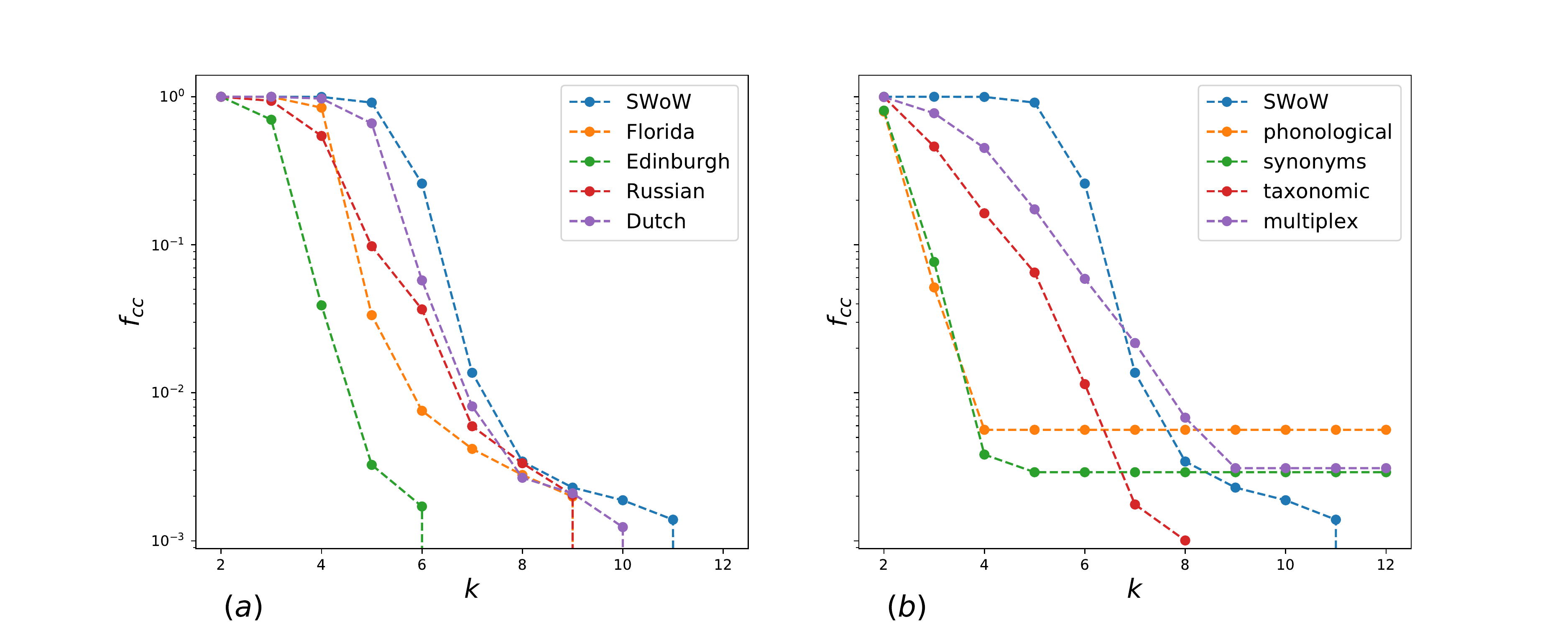}}
\caption{(a) The size of k-clique percolation cluster in dependence on the value k for different free association datasets. (b) The size of k-clique percolation cluster in dependence on the value k for different English semantic networks.}
\label{fig01}
\end{figure}

We calculated the fraction of nodes $f_{cc}$, included in k-clique percolation cluster for different values k. The dependencies for different free association datasets are presented in Fig.\ref{fig01}(a). 
Firstly, note that almost all words are included in 3-clique percolation cluster, the existence of this cluster explains the high transitivity and average clustering coefficient, see  Table ~\ref{tab:1}. Secondarily,all free association  networks, except   Edinburgh dataset demonstrates k-clique percolation for large k, i.e., the clusters of $k=5$ and $k=6$ contain essential fraction of words and for SWoW-EN dataset - almost all words. We suggest, that this feature is result of higher density of SWoW-EN dataset, which is achieved by including weak associations.  A more detailed analysis of percolation depending on the association strength is presented in the Section 5.

In Fig.\ref{fig01}(b) the dependencies for semantic networks of different nature are presented.In contrast to the networks of free associations, phonological and synonymous networks form a 3-clique percolation cluster only partially, and clusters of higher orders are completely absent, despite the fact that these networks are characterized by higher values of transitivity and clustering.
We also calculated the respective dependence for the so-called multiplex network, in which we considered three layers:phonological, taxonomic and synonyms. For such a network, we observe clusters of the order of 4 and 5. Thus, we can assume that the variety of links, weak association in SWoW-EN, or different types of links in the multiplex network ensure the existence of high-order clique clusters.

\newpage
\section{Structural features and clustering in SWOW-EN network}
\begin{figure}[ht]
\centerline{\includegraphics[width=20cm]{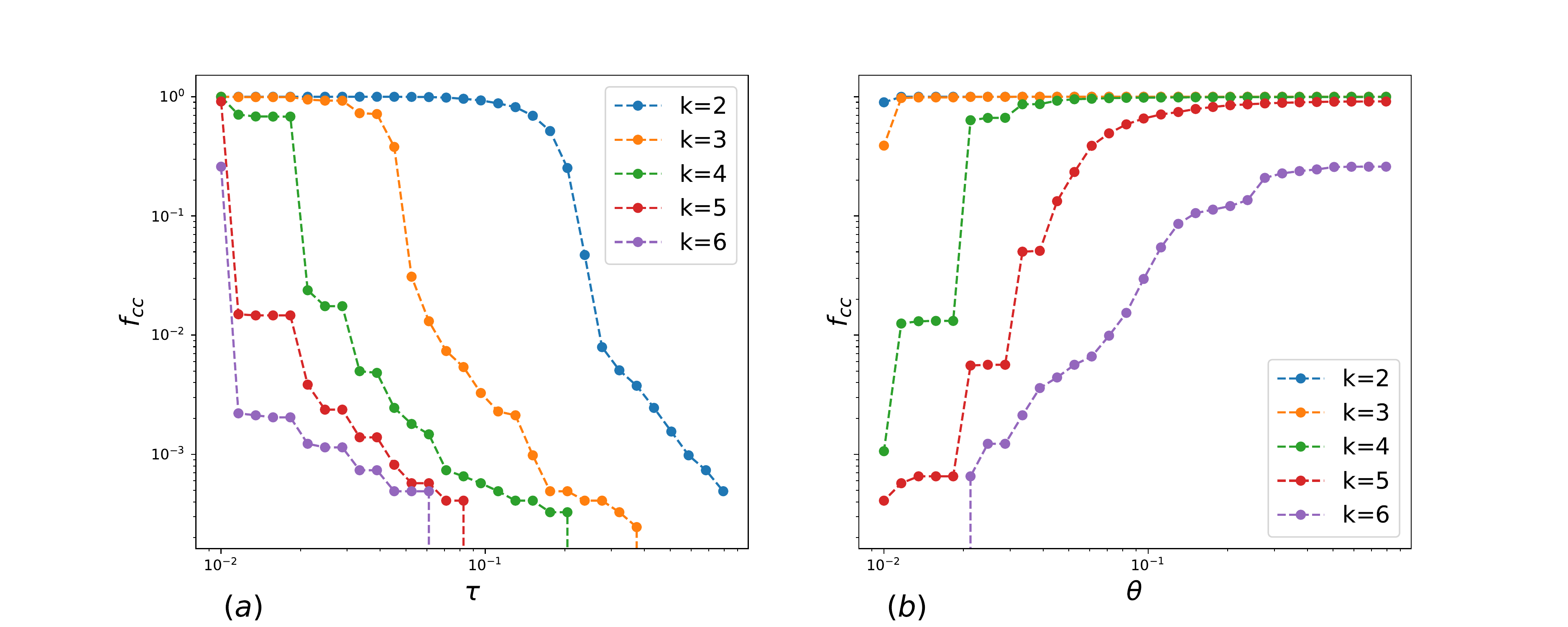}}
\caption{(a) The size of k-clique percolation cluster in dependence on the threshold $\tau$ for different values k in SWOW-EN. (b) The size of k-clique percolation cluster in dependence on the threshold $\theta$ for different values k in SWOW-EN.}
\label{fig02}
\end{figure}

\begin{figure}[ht]
\centerline{\includegraphics[width=12cm]{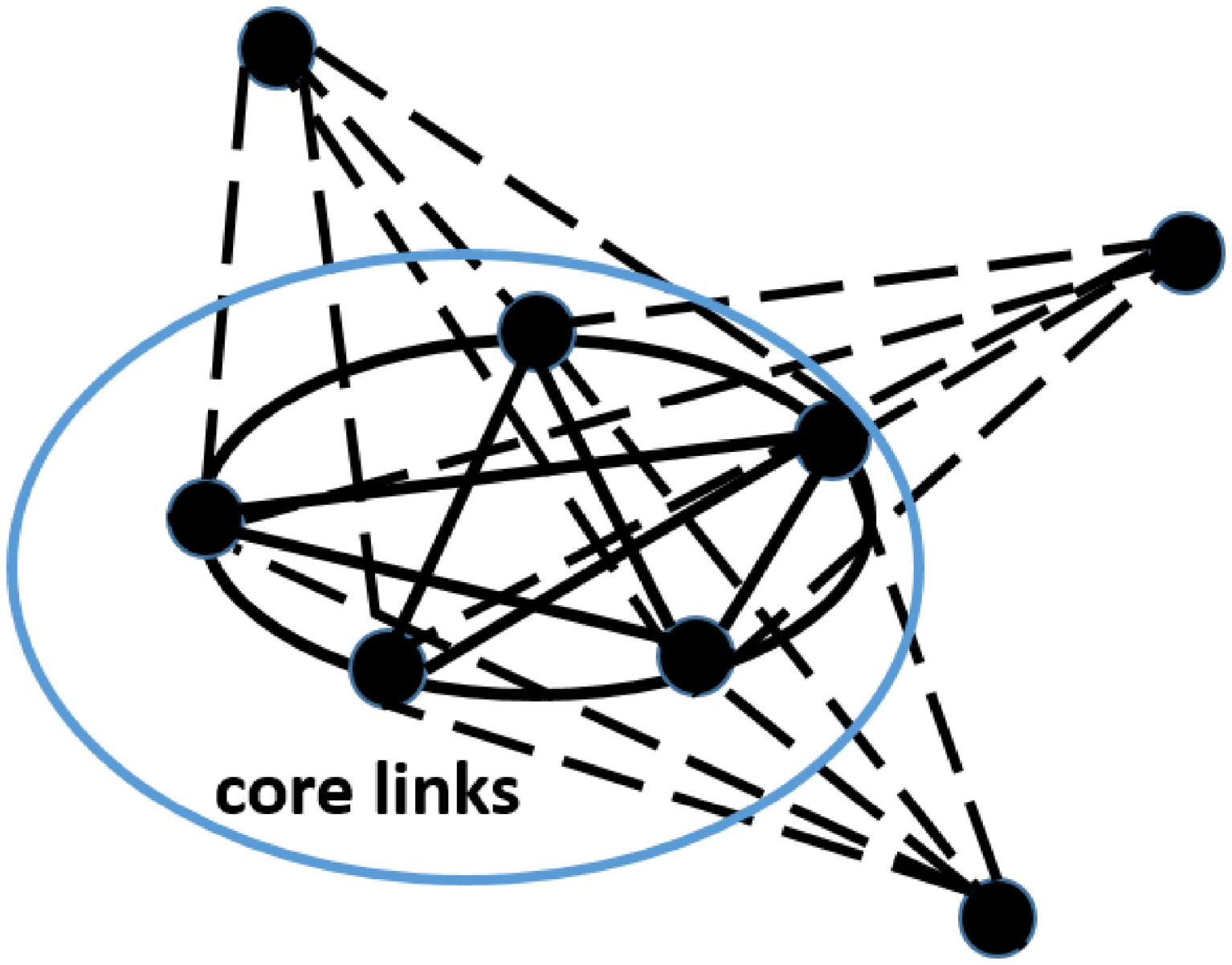}}
\caption{(a) Three k-cliques are adjacent ($k=6$) through the central (k-1)-clique, which could be considered as a "core".}
\label{fig03_2}
\end{figure}

We analyze k-clique community clusters in dependence on the association strength. For this aim we perform following numerical experiments. 
In first simulation we take a threshold $\tau$ for association strength and delete all links of weights \emph{less} than the threshold. Fig.\ref{fig02}(a) presents the fraction of nodes including in K-clique community cluster of $k=2,3..6$ in dependence on the threshold $\tau$. 
We observe, that k-clique community clusters of higher order ($k=5$ and $k=6$) exist only for initial network state and almost disappear at a small threshold. Percolation clusters for $k = 3$ and $k=4$ include all vertices up to sufficiently high threshold $\tau$, indicating the stability of network community organisation. 
The second simulation is following. We establish a threshold $\theta$ for association strength and delete all links of weights \emph{more or equal} than the value $\theta$, i.e. we analyze a subgraph of weak associations.  The respective dependencies for different k are depicted in Fig.\ref{fig02}(b). 3-clique percolation cluster is not sensitive to the threshold $\theta$ and exists for all weak subgraphs.
The percolation clusters  for $k = 4$ and $k=5$ include all words for high threshold and abruptly decrease at small values $\theta$. Interestingly, that for  $k=6$ even for very high $\theta$ the percolation cluster contains only some part of nodes. This result shows that strong free associations can be considered as "core" links, which are involved in few cliques, providing intersection of cliques in percolation cluster. While weak associations form rather  a "shell" of clique community, see Fig.\ref{fig03_2}. 

\begin{figure}[ht]
\centerline{\includegraphics[width=20cm]{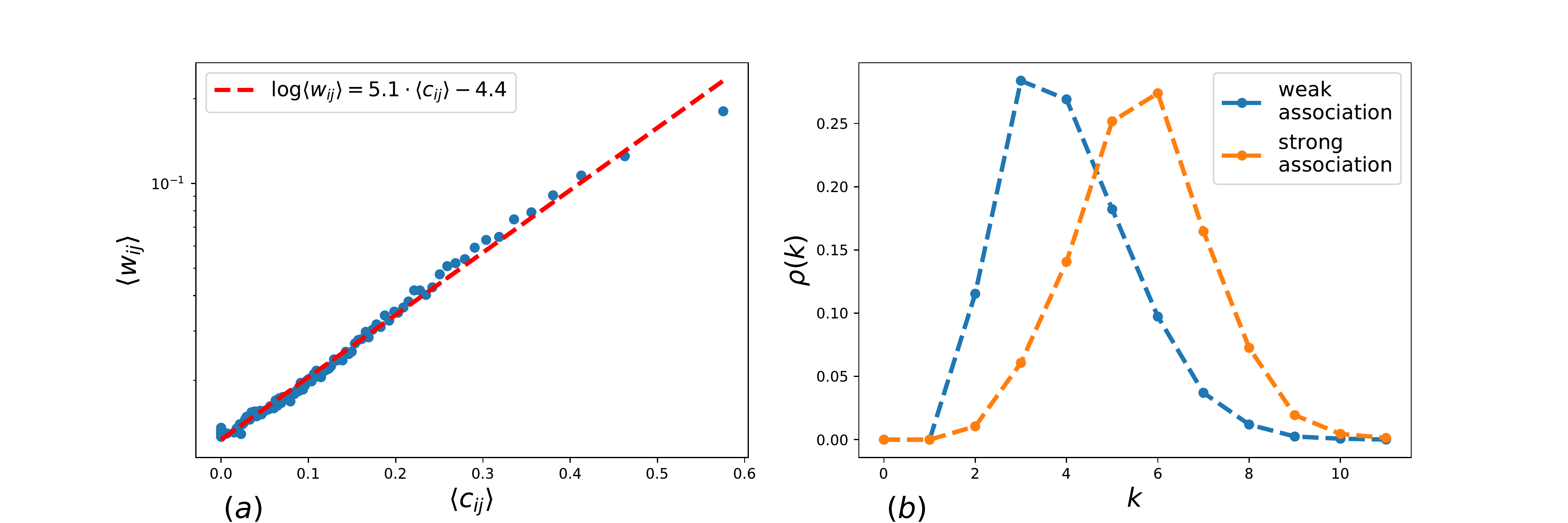}}
\caption{(a) The dependence of average association strengths on the edge clustering. (b) The  distribution of maximal clique sizes for weak and strong associations.}
\label{fig03}
\end{figure}

This assumption is confirmed by the study of the distribution of triangles belonging to the links depending on association strength. We introduce an \emph{edge clustering coefficient} for as follows:
$$
C_{ij}=\frac{N_T(ij)}{\min(k_i, k_j)-1},
$$
where $N_T(ij)$ is the number of triangles, containing the edge~$(i,j)$, $k_i, k_j$ are the degrees of $i$ and $j$ nodes respectively.  Like a clustering coefficient of a node, the value $C_{ij}$ shows the fraction of triangles and lies in the range $\left[0,1\right]$. Note that the introduced clustering coefficient $C_{ij}$ correlates with a topological overlap for nodes $i$ and $j$ in case of their adjacency \cite{ravasz2002}. 
We found the clustering coefficient for each edge in the free association network SWOW-EN, sorted them and splitted into 
$ n = 100 $ intervals of equal size. For each intervals $l, l=1,2,\dots,n$ we calculated the average values for the clustering coefficient $\left<c_{ij}^l\right>$ and  for the association strengths $\left<w_{ij}^l\right>$. Fig.\ref{fig03}(a) presents the dependence of the average association strength on the respective clustering coefficient in given interval. One can see that the dependence is fitted by the curve $\log\left<w_{ij}\right>=5.1\cdot\left<c_{ij}\right>-4.4$. Thus, we observe a positive correlation between the number of triangles, belonging to a link and its association strength. Note that this correlation was not discussed before and it is interesting by itself and can be used in modeling the human lexicon. 
 Besides, we introduce \emph{k-clique number} of an edge as the maximal clique size, containing the edge. In Fig.\ref{fig03}(b) the distributions of k-clique numbers are presented for the weakest association links, i.e. $w_{ij}=0.01$ and for the strongest association links, $w_{ij}>0.1$. 
 
 \section{Simulation of clique organization in free association networks}
 Network models of language structure are discussed in  \cite{arenas2010,dorogovtsev2001}. Particularly, Dorogovtsev and Mendes \cite{dorogovtsev2001} proposed a stochastic theory of the evolution of human language, which treats language as a self–organizing network of interacting words. It is well known that language evolves, then the question is what kind of growth (in the sense of increase of lexical repertoire) leads to a self-organized structure with characteristic scale-free degree distribution. Dorogovtsev and Mendes’ scheme of the language network growth is following. A new word is connected to some old one $i$ with the probability proportional to its degree $k_i$ (Barabasi and Albert’s preferential attachment); additionally, at each time step,$c$ new edges randomly emerge between old words, where $c$ is a constant coefficient that characterizes a particular network. 
This model explained very well power law degree distribution and small world properties of semantic networks. 

\begin{figure}[ht]
\centerline{\includegraphics[width=12cm]{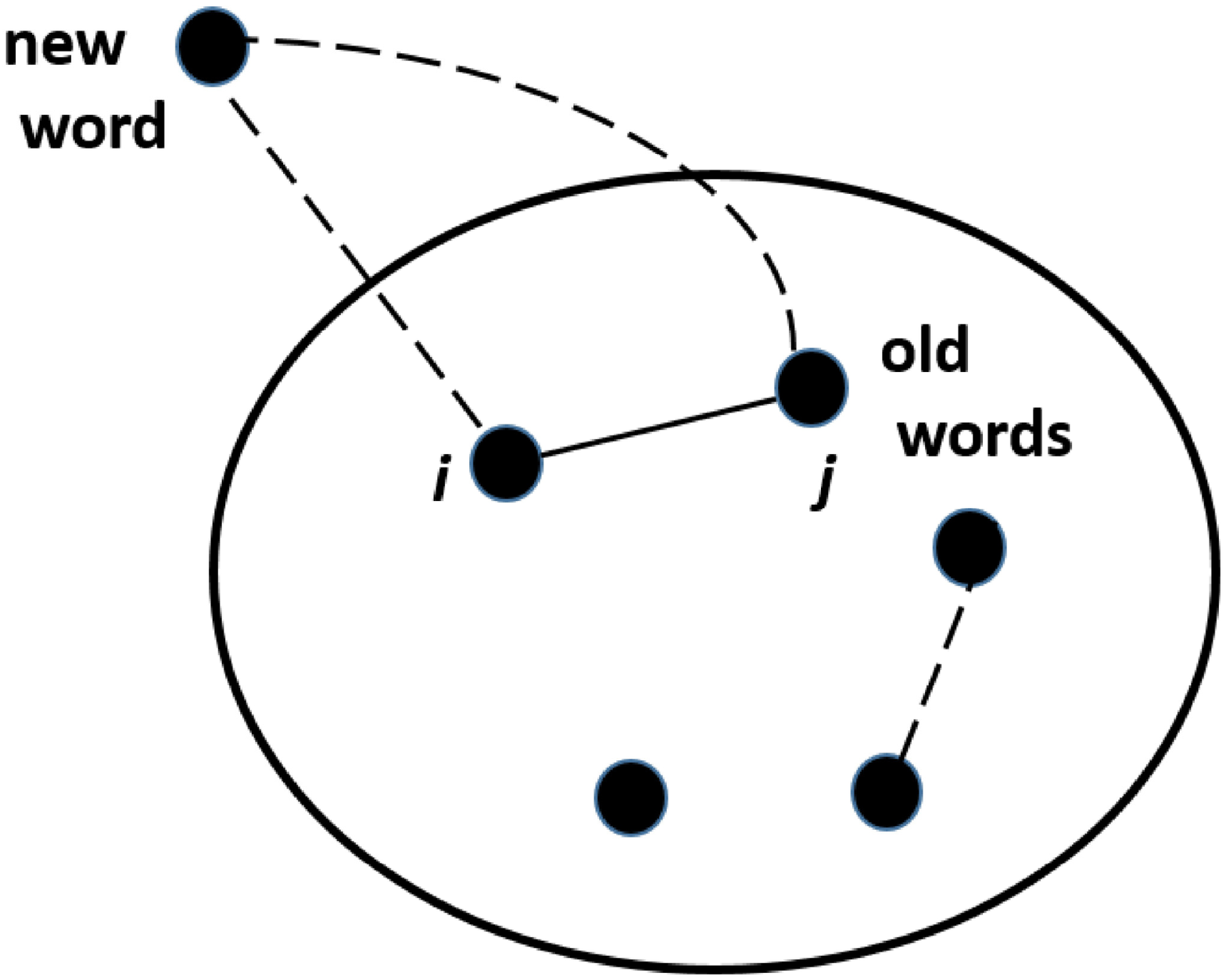}}
\caption{Network model description: a new word is connected to a link $(i,j)$ by preferential attachment; in addition, random links between old words emerge. Existing links are depicted by solid line, new links are dashed.}
\label{fig04}
\end{figure}

To describe clique organization in semantic networks we propose a new model based on Dorogovtsev and Mendes’ mechanism, presented in Fig.\ref{fig04}(a). In our model a new word is connected to $2m$ existing \emph{linked} words $i$ and $j$ with the probability proportional to the sum degree $k_i+k_j$, forming a triangular; in addition, at each time step, we add $c$ new edges randomly between old words. The network evolution begins with an initial small  Erdos–Renyi random graph $G(l,p_0)$.

\begin{figure}[ht]
\centerline{\includegraphics[width=20cm]{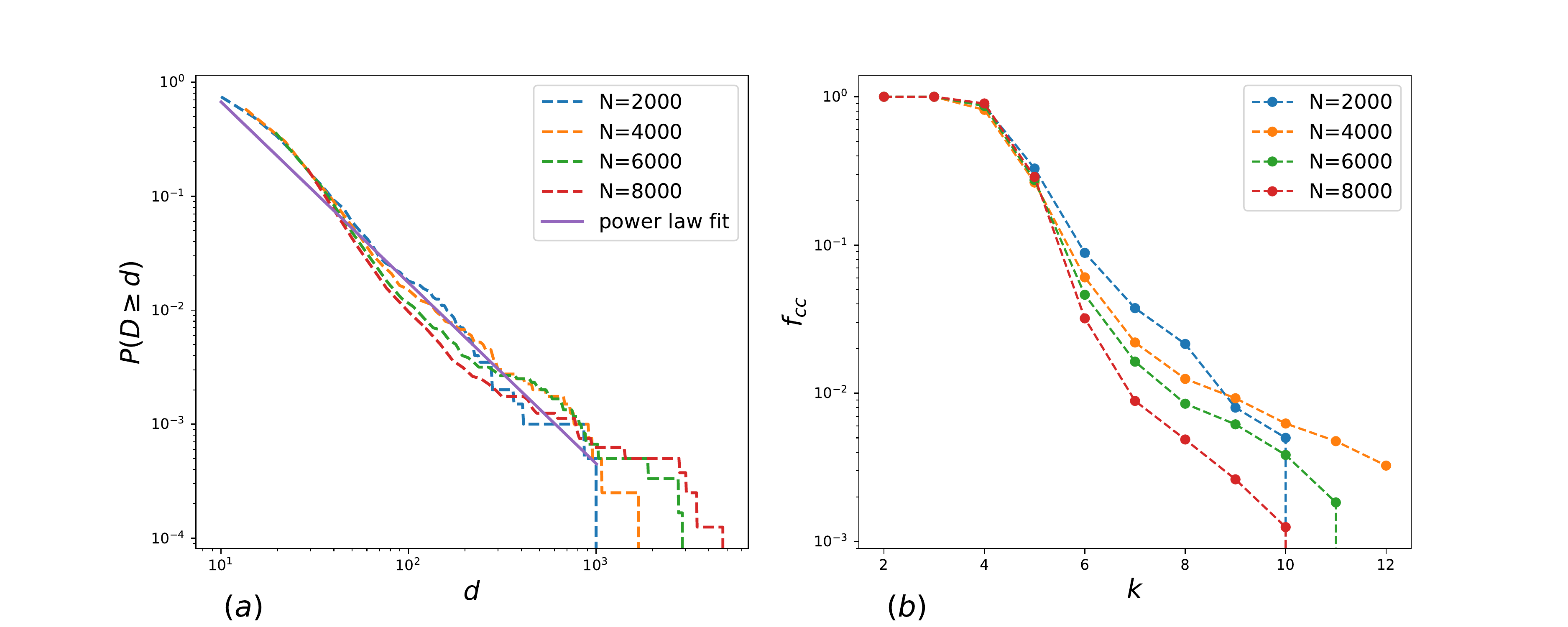}}
\caption{(a) Complementary cumulative degree distribution function for simulated networks of different sizes; (b) The size of k-clique percolation cluster in dependence on the value k for networks of different sizes.}
\label{fig05}
\end{figure}

\begin{table}[ht]
\caption{Structural properties of simulated networks.}
\begin{center}
\begin{tabular}{l|l|l|l|l|l|l}
\hline \hline
 Nodes\hspace{1cm} & Edges & Density & Transitivity & Clustering &$p_c(2)$& $p_c(3)$ \\ \hline\hline
 2~000& 23~213&0.0116 &0.048&0.175&$5\cdot 10^{-4}$& 0.0158\\ \hline
 4~000&46~783&0.0058 &0.028&0.158&$2.5\cdot10^{-4}$&0.0111 \\ \hline
6~000&69~307&0.0039&0.016&0.172&$1.66\cdot10^{-4}$&0.0091\\ \hline \hline
8~000&91~275&0.0028&0.010&0.187&$1.25\cdot10^{-4}$&0.0079
\\ \hline \hline
\end{tabular}
\end{center}
\label{tab:2}
\end{table}
We simulated the networks of different sizes with the model parameters $m=4$, $c=4, l=20, p_0=0.1$. Structural properties of the networks are summarized in Table ~\ref{tab:2}. All networks are sparse with the density, $p_c(2)<\rho<p_c(3)$, and high clustered.
 Degree distributions of the networks are fitted by power law $p=Cd^{-\gamma}$ with $\gamma=2.6$, see Fig.\ref{fig05}(a). The size of  k-clique percolation cluster in dependence on the value $k$ demonstrates the same behaviour as observed for free association networks (Fig.\ref{fig05}(b)) and does not depend on the network size. Thus, the assumption of preferential attachment to an edge rather than a single word may explain clique organization in free association networks. 

 \section{Towards the explanation of Miller's $7 \pm 2$ rule ?}
 
 In this Section we shall make the conjecture concerning more general meaning 
 of our findings. It was remarked long time ago \cite{miller} that 
 many phenomena concerning the consuming  information by the 
 human brain for a short period of time have a natural restriction of the number of  
 controlled items. This number is estimated by the Miller's 
 $7 \pm 2$ rule which implies the restricted ability of brain 
 to handle with the information processing. There were a few attempts
 to apply the underlying network structure to explain the Miller rule
 for the limit of capacity of working memory \cite{cowan}.
 
 From the phychological viewpoint  three groups of 
 mechanisms behind the limit capacity have been suggested (see, \cite{3mech} for
 the review). First approach implies the short time decay of the groups
 of items because of some reason. Second mechanism of limited resource claims that there are no enough resources for higher capacity.
A resource is considered as some limited quantity
that enables a cognitive function or process.
According to the third mechanism 
our ability to hold several representations available at the same time is
limited by mutual destructive interference between these representations.
 As an example of this mechanism one could have in mind the interference
 of frequency bands in brain activity. Indeed it is known that a few bands 
 are simultaneously involved
 in the processing of working memory. None of these mechanisms can
 be considered as fully satisfactory. Another network motivated
 approach  \cite{glassman} utilizes the mathematical result concerning
 the plane colouring by four colors. This idea was conjectured  to be relevant
 for the smaller critical number of items discussed in \cite{cowan}.
 
 Can we gain some new insight concerning the mechanism behind the  Miller's rule from our study?
 Let us  assume that the free association
 tests are the specific probes of the working memory. This 
 assumption has been discussed in the literature before 
 ( see, for instance \cite{goni})
 and it is natural since the time allowed for the 
 performance of tests is quite restricted. Hence let us assume 
 that  networks of free associations carry the information
 about the working memory and their structure
 code the information  concerning the group of related
 stimuli during the test. We conjecture that
 keeping the information about the linked group of k stimuli is 
 encoded in the k-clique percolation. Hence our finding that the
 k-clique percolation for the SWOW-EN and Dutch networks is possible 
 only for $k \le 6$
 can be interpreted as the example of the Miller's rule. 
 
 One may concern 
 that in the working memory setup we have "percolation in time"
 to keep the group of stimuli  as a whole for some 
 period of time. On the other hand in the clique percolation we have a "percolation 
 on the network" keeping the clique intact when moving along the
 graph. However to some extent the formation of the association network
 can be considered as the growing network model.
 This viewpoint if true suggests the new  perspective of explanation
 of the Miller's rule for all behavioural situations when the 
 network description is available. One has to estimate the maximum
 size of percolating clique in the particular network architecture.
 It provides the answer for the capacity of the working memory
 which ensure the propagation of the linked group of items 
 in time. 
 
 One could question why only the low size clique percolation
 in the human brain is possible although naively we could expect
 that the brain would prefer the higher working memory 
 capacity. In particular 
 a limit of the working memory capacity for human is higher then 
 of other animals \cite{animals} and it is assumed that higher
 working memory corresponds to higher intellect. The answer
 certainly should involve some evolutionary arguments and
 at least two alternative scenario are possible. First,
 presumably the network architecture admitting higher clique percolation 
 is in contradiction with some other vital properties of a brain
 encoded in connectome architecture. 
 
 Secondly, we could assume that the particular evolutionary rules 
 (see,\cite{evolution} for the review) for the 
 corresponding network dynamically bring it to the 
 particular capacity limit somewhat in the spirit 
 of self-organized criticality. In the previous Section
 we have supported this possibility suggesting 
 the non-conventional version of the preferential
 attachment procedure  which indeed yields the reasonable 
 value of $k_c=(5-6)$.
 
 \section{Conclusion}
 
 In this paper we have analysed the k-clique percolation 
 in the free association networks and semantic networks for few languages. 
 There are few findings of our study which seem to be important.
 First, using the  traditional  approach we have
 investigated the structural network properties 
 via the percolation theory and made a few 
 new observations 
 \begin{itemize}
     \item There is the critical value $k_c$ 
 for the maximal size of a percolating k-clique 
 for the SWOW-EN network and semantic networks.  The higher
 clique with $k>k_c$ can not percolate through these networks
 \item The density of the analyzed network does not allow
 the percolation of $k>2$ cliques if the network is considered
 as random. This means that our study confirms the non-randomness
 of the free association networks
 \item Imposing the thresholds on the link weights we investigated
 the role of weak and strong associations on the k-clique size and
 percolation. It is found that the strong associations play 
 the key role in the k-clique percolating cluster while the weak
 associations provide a kind of shadow which however 
 is necessary ingredient supporting the observations
 made in \cite{valba2021}. The clear-cut dependence
 between the averaged local weights and the local connectivity has
 been found.
 \end{itemize}
 
 Secondly we assume that our findings provide the
 additional information not only on the structure
 but also for the processing on the network. 
 Namely the free association networks
 can be considered as the peculiar  probe of
 the working memory. Therefore the critical $k_c$ for the 
   k-clique percolation we have found  presumably can be interpreted
   as the limitation of capacity of the working memory
   and therefore can be new nontrivial 
   qualitative mechanism behind the
   Miller's law. This can be further checked by investigating
   a threshold in k-clique percolation for other 
   cognitive networks involving short-time performances
   probing working memory.
   
   It would be interesting to elaborate further the possible
   origin of $k_c$ actual value. Presumably it can
   be established evolutionary \cite{evolution} as a optimal result of 
   competition between the clique percolation related to the
   working memory and another properties of the 
   connectome responsible for  important cognitive
   properties. Another possibility demonstrated in our study
   is that the specific version of the preferential attachment
   evolution mechanism  yields the critical value $k_c=(5-6)$
   via a kind of self-organized criticalily. It would be interesting
   to test our new evolutionary rule in the other cognitive
   processes.
   
 We are grateful to K. Anokhin for the useful comments.
 The work was supported in part by grant N 075-15-2020-801 by
  Ministry of Science and Higher Education of Russian Federation (A.G), 
  and grant  RFBR  18-29-03167 (A.G.,O.V.)



\begin{thebibliography}{99}
\bibitem{siew2019} C. S. Q. Siew, D. U. Wulff,  N. Beckage, and Y. Kenett,   Cognitive Network Science: A review of research on cognition through the lens of network representations, processes, and dynamics. Complexity, 1-24 (2019). 

\bibitem{rev2}
A. Baronchelli, R. Ferrer,C. R. Pastor-Satorras, N. 
Chater and M. H. Christiansen, 
Networks	in	Cognitive	Science,
arxiv 1304.6736.

\bibitem{stella2018} M. Stella, N.M. Beckage, M. Brede, and M. De Domenico, Multiplex model of mental lexicon reveals explosive learning in humans, Scientific Reports, {\bf 8}, 2259 (2018).

\bibitem{stella2019} M. Stella Modelling Early Word Acquisition through Multiplex Lexical Networks and Machine Learning, Big Data Cogn. Comput., {\bf 3}, 10 (2019).

\bibitem{kenett2014} Y. N. Kenett, D. Anaki and M. Faust,  Investigating the structure of semantic networks in low and high creative persons. Frontiers in Human Neuroscience, {\bf 8 (407) }, 1–16 (2014).

\bibitem{smith2013} K.A. Smith., D.E. Huber, and E. Vul,  Multiply-constrained semantic search in the Remote Associates Test, Cognition, {\bf 128}, 64 (2013).

\bibitem{bourgin2014} D. D. Bourgin, J.T. Abbot, and T.L. Griffiths,  Empirical Evidence for Markov Chain Monte Carlo in Memory Search, Proceeding of the Annual Meeting of the Cognitive Science Society, {\bf 36}, 224 (2014).

\bibitem{olteteanu2015} A.-M. Olteteanu and Z. Falomir,  ComRAT-C: A Computational Compound Remote Associates Test Solver based on Language Data and its Comparison to Human Performance, Pattern Recognition Letters, {\bf 67}, 81 (2015).

\bibitem{olteteanu2017} A.-M. Olteteanu and H. Schultheis, What determines creative association? Revealing two factors which separately influence the creative process when solving the remote associates test, Journal of Creative Behavior, {\bf 53}, 389 (2017).

\bibitem{valba2021} O. Valba, A. Gorsky, S. Nechaev, M. Tamm, Analysis of English free association network reveals mechanisms of efficient solution of Remote Association Tests. PLOS ONE {\bf 16(4)}: e0248986 (2021).

\bibitem{mednik1962} S. Mednick, The associative basis of the creative process, Psychological Review, {\bf 69}, 220 (1962).

\bibitem{dedeyne2019} S. De Deyne, D. J. Navarro, A. Perfors, M. Brysbaert, and G. Storms, The “Small World of Words” English word association norms for over 12,000 cue words, Behavior Research Methods, {\bf 51}, 987-1006 (2019).

\bibitem{kenett2018} Y. N. Kenett,  O. Levy, D.Y. Kenett, H.E. Stanley, M. Faust and S. Havlin, Flexibility of thought in high creative individuals represented by percolation analysis. Proceedings of the National Academy of Sciences, 201717362 (2018).

\bibitem{stella2020} M. Stella,
Multiplex networks quantify robustness of the mental lexicon to catastrophic concept failures, aphasic degradation and ageing,
Physica A: Statistical Mechanics and its Applications, 554 (2020), Article 124382

\bibitem{arenas2011}
J. Borge-Holthoefer, Y. Moreno, A. Arenas;
Modeling abnormal priming in Alzheimer’s patients with a free association network
PLoS One, 6 (8) (2011), Article e22651

\bibitem{derenyi2005} I. Derenyi, G. Palla and T. Vicsek, Clique percolation in random networks, Phys.Rev. Lett., 94 (2005), 160202.

\bibitem{palla2007} G. Palla, I. Derenyi and T. Vicsek, The critical point of k-clique percolation in theErdos–Renyi graph, J. Stat. Phys., 128 (2007), 219–227.

\bibitem{kenett21}
A. L.Cosgrove,Y. N.Kenett, R. E.Beaty,M. T.Diaz;
Quantifying flexibility in thought: The resiliency of semantic networks differs across the lifespan,
Cognition, 211, 2021, 104631

\bibitem{dis1} P. Hoffman, J.L. McClelland and M.A. Lambon-Ralph,
Concepts, control, and context: A connectionist account of normal and disordered semantic cognition,
Psychological Review, 125 (3) (2018), pp. 293-328

\bibitem{dis2} T.T. Rogers,K. Patterson, E. Jefferies, M.A. Lambone-Ralph
Disorders of representation and control in semantic cognition: Effects of familiarity, typicality, and specificity
Neuropsychologia, 76 (2015), pp. 220-239

\bibitem{miller}
G. Miller, The magical number seven,plus or minus two. Some limits on our capacity for processing information, Psichol.Rev. 63, 81-97,1956

\bibitem{cowan}
N. Cowan, The magical number 4 in short-term memory: A reconsideration of mental storage capacity. Behavioral and Brain Sciences, 24, 87-185,2001 \\
N. Cowan, The magical mystery four: How is working memory Capacity Limited and Why?,
Curr.Dir.Psychol Sci. 19,51-57, 2010

\bibitem{nelson2004} D.L. Nelson, C.L. McEvoy, and T.A. Schreiber, The University of South Florida free association, rhyme, and word fragment norms, Behavior Research Methods, Instruments, \& Computers, {\bf36}, 402-407, (2004).

\bibitem{edinburgh} Coltheart, M. The MRC psycholinguistic database. The Quarterly Journal of Experimental Psychology 33, 497–505 (1981).

\bibitem{worddata} WolframResearch. WordData source information.

http://reference.wolfram.com/language/note/WordDataSourceInformation.html.

\bibitem{miler} Miller, G. A. WordNet: a lexical database for English. Communications of the ACM 38, 39–41 (1995).

\bibitem{rus_th} Russian associative dictionary. An associative thesaurus of the modern Russian language. In 3 parts, 6 books / Yu.N. Karaulov, Yu.A. Sorokin, EF Tarasov, N.V. Ufimtseva, G.A. Cherkasova. Book. 1, 3, 5. Direct vocabulary: from stimulus to reaction. Book 2, 4, 6. Reverse vocabulary: from reaction to stimulus. M., 1994, 1996, 1998.

\bibitem{swow} Dutch Data, https://smallworldofwords.org/


\bibitem{ravasz2002} Ravasz E, Somera AL, Mongru DA, Oltvai ZN, Barabási AL. Hierarchical organization of modularity in metabolic networks. Science. (2002), 297(5586):1551-5.

\bibitem{arenas2010} Borge-Holthoefer, J., Arenas, A. (2010)Semantic Networks: Structure and Dynamics, Entropy,12(5), 1264-1302, doi:10.3390/e12051264

\bibitem{dorogovtsev2001} Dorogovtsev, S.; Mendes, J. Language as an evolving word web. Proc. R. Soc. Lond., B, Biol. Sci.
2001, 268, 2603–2606.

\bibitem{3mech} K. Oberauer,S. Farrell,C. Jarrold,S. Lewandowsky
"What Limits Working Memory Capacity?"
Psychological Bulletin,
2016, Vol. 142, No. 7, 758–799

\bibitem{glassman} R. B. Glassman,"Topology and graph theory applied to cortical anatomy may help explain working memory capacity for three or four simultaneous items", Brain Research Bulletin 60 (2003) 25–42
\bibitem{goni}	Goñi,	J.,	et	al. (2010)	Switcher-random-walks:	a	cognitive-inspired	
mechanism	for	network	exploration.	Int.	J.	Bifurcation	Chaos 20,	913-922

\bibitem{animals} L. A. Hahn, D. Balakhonov, E. Fongaro
, A. Nieder, J. Rose. 
"Working memory capacity of crows and monkeys 
arises from similar neuronal computations"
bioRxiv preprint doi: https://doi.org/10.1101/2021.08.17.456603



\bibitem{evolution}
Majid Manoochehri,
"Up to the magical number seven: An evolutionary perspective on the capacity of short term memory"
Heliyon,v. 7,  5, E06955, 2021










\end{thebibliography}
\end{document}